\documentclass[10pt]{elsarticle}

\usepackage{amsmath}
\usepackage{amssymb}

\biboptions{sort&compress}

\newtheorem{remark}{Remark}

\journal{Proceedings of the Conference ``Symmetry~2024'', 
Nakhon Ratchasima, Thailand.}

\allowdisplaybreaks

\begin{document}

\begin{frontmatter}


\title{Exact solutions and automorphic systems of the \\ geopotential forecast equation}



\author[address]{Evgeniy~I. Kaptsov\corref{correspondingauthor}}

\ead{evgkaptsov@math.sut.ac.th}

\affiliation[address]{organization={School of Mathematics, Institute of Science, Suranaree University of Technology},
postcode={30000},
city={Nakhon Ratchasima},
country={Thailand}}

\cortext[correspondingauthor]{Corresponding author}



\begin{abstract}
The study of the recently constructed group foliation for the geopotential forecast equation is continued. 
The group foliation consists of two systems, namely the automorphic and resolving systems, the analysis of which facilitates the derivation of invariant solutions for the original equation.
As obtaining a general solution to the resolving system (even to its reductions on subgroups) is problematic, 
its various particular solutions are considered. 
Consequently, the question arises concerning the specific forms of automorphic systems that correspond to exact solutions obtained 
through alternative methods. 
This is of interest for both comparing solutions derived through different approaches and for the integration of specific automorphic systems.
The problem is discussed in a number of examples.
\end{abstract}






\end{frontmatter}

%
%

\section{Introduction}


The prediction of geopotential in regions characterized by rotating air masses at intermediate altitudes within the atmosphere
relies on the geopotential forecast equation\footnote{The equation is also referred as barotropic vorticity equation on the beta-plane.} 
(GFE)~\cite{KaptsovEI:bk:Kibel1957,KaptsovEI:bk:HandbookLie_v2}
\begin{equation} \label{KaptsovEI:EqMain0}
(H_{xx} + H_{yy})_t - H_y (H_{xx} + H_{yy})_x + H_x (H_{xx} + H_{yy})_y + \beta H_x = 0.
\end{equation}
Here $t$ is time, $x$ and $y$ are coordinates, $(-H_y, H_x)$ is the two-dimensional velocity potential,
and the constant~$\beta$ is related to the $\beta$-approximation of the Coriolis parameter~\cite{KaptsovEI:bk:WallaceHobbs[2006]}.

Exact solutions of~(\ref{KaptsovEI:EqMain0}) have been studied in various 
works~\cite{KaptsovEI:bk:Katkov_Geostr[1965],KaptsovEI:bk:Katkov_Geostr[1966],KaptsovEI:bk:HandbookLie_v2} since the 1960s. 
An intensive study of symmetries and invariant solutions of the~GFE
and its generalizations have been carried out in a series 
of publications~\cite{KaptsovEI:Bihlo_2009a,KaptsovEI:Bihlo_2009b,KaptsovEI:bk:BihloPop_Geostr_InvSols[2011],KaptsovEI:bk:BihloPop_Geostr_numeric[2019]}, 
as well as in several separate works of Chinese scientists~\cite{KaptsovEI:Fei_2004, KaptsovEI:TangXiao-Yan_2008, KaptsovEI:bk:xu2008algebraic}. 
Conservation laws of the~GFE have been studied in~\cite{KaptsovEI:bk:Sameerah2020} and~\cite{KaptsovEI:bk:Kaptsov2023GroupFol}.

In~\cite{KaptsovEI:bk:Kaptsov2023GroupFol}, the author investigated group foliations of the~GFE, complementing previously known results in the symmetry analysis of~(\ref{KaptsovEI:EqMain0}). 
The group foliations~(stratifications) approach~\cite{KaptsovEI:bk:Ovsiannikov1978} is based on the theory of differential invariants and involves transitioning from the study of the original equation~(or system of equations) to the analysis of two systems:~automorphic and resolving. 
Sometimes a problem formulated in this manner proves to be simpler than the original one~\cite{KaptsovEI:bk:Ovsyannikov_strat[1970],KaptsovEI:bk:Ovsiannikov1978}. 
Group foliations are particularly advantageous when a Lie algebra admitted by equations includes 
a generator corresponding to an infinite-dimensional group of transformations, a situation typical, for instance, 
for hydrodynamic-type equations in Lagrangian coordinates~\cite{KaptsovEI:bk:DorKapMel_SW2D,KaptsovEI:bk:MelKapGasDynFolMDPI2024}. 
In such cases, the resolving system admits a finite-dimensional Lie algebra, facilitating a more straightforward analysis. 
Furthermore, invariant solutions derived from reductions of the resolving system may have a more general form than those 
obtained through standard methods~\cite{KaptsovEI:bk:Golovin_strat[2003]}.

As it is often problematic to obtain a general solution of a resolving system, one has to be content with its particular solutions. 
Each particular solution of a resolving system distinguishes a specific automorphic system and, then, 
a specific equivalence class of invariant solutions of the original system. 
In this regard, the inverse problem arises:~knowing an invariant solution, construct a corresponding automorphic system from it. 
This may be of interest, for example, when analyzing exact solutions obtained by other methods. 
Whether such an exact solution can be derived from a group foliation and, if so, which automorphic system it corresponds to? 

The problem of constructing automorphic systems from exact solutions is discussed in~\cite{KaptsovEI:bk:Ovsiannikov1978}. 
There it is shown that for a given invariant solution~$s$ there always exists a corresponding automorphic system.
This is achieved due to the fact that, starting from some~$k \geqslant 0$, the~$k$th prolongation of the orbit of~$s$ 
has a projection in the space of differential invariants of order~$k$ and can be specified by a system of equations 
relating differential invariants of order less or equal to~$k$. 
Such a system is automorphic, as the projection of the orbit of any of its solutions $\widetilde{s}$ 
can be embedded in the projection of the orbit of the solution~$s$.

Unfortunately, the theoretical justification proposed in~\cite{KaptsovEI:bk:Ovsiannikov1978} is not accompanied by illustrative examples. 
Similarly, in~\cite{KaptsovEI:bk:Kaptsov2023GroupFol}, while the problem of constructing automorphic systems from exact solutions 
for the~GFE was outlined, detailed examples were not provided in the discussion. 
In practice, demonstrating that a given exact solution satisfies an automorphic system may require considerable effort. 
The present publication is devoted to the consideration of specific examples elaborated as a result 
of the analysis of the~GFE solutions given in~\cite{KaptsovEI:Bihlo_2009a} and~\cite{KaptsovEI:bk:HandbookLie_v2}.

\smallskip

The paper is organized as follows.
In Section~\ref{KaptsovEI:sec:Sec2_theory}, the admitted Lie algebra and the group foliation (case $\beta \neq 0$) for the~GFE are given. 
The main results are presented in Section~\ref{KaptsovEI:sec:Sec3_Results}  
devoted to the construction of automorphic systems corresponding to exact solutions.
Finally, Section~\ref{KaptsovEI:sec:Sec4_Concl} provides concluding remarks.

\section{Group Foliation of the GFE (Case $\beta \neq 0$)}
\label{KaptsovEI:sec:Sec2_theory}

As the details of the theory of differential invariants and group foliations are beyond the scope of this presentation, 
the interested reader may refer to~\cite{KaptsovEI:bk:Ovsiannikov1978}. 
This section primarily presents the results from~\cite{KaptsovEI:bk:Kaptsov2023GroupFol}, which are essential for the subsequent discussion. 
We also stay focused only on the case $\beta \neq 0$, 
as the group foliation for~$\beta=0$ is overcomplicated and turns out to be practically unreasonable~\cite{KaptsovEI:bk:Kaptsov2023GroupFol}.

\smallskip

In case $\beta \neq 0$, the Lie algebra admitted by~(\ref{KaptsovEI:EqMain0}) consist of the following generators~\cite{KaptsovEI:bk:HandbookLie_v2}
\begin{equation} \label{KaptsovEI:alg}
\def\arraystretch{1.5}
\begin{array}{c}
X_1 = \partial_t,
\qquad
X_2 = \partial_y,
\qquad
X_3 = -t \partial_t + x \partial_x + y \partial_y + 3 H \partial_H,
\\
X_{\infty} = f \partial_x + (g - y f^\prime) \partial_H,
\end{array}
\end{equation}
where $f$ and $g$ are arbitrary functions of $t$. 
To construct a group foliation we also require $f$ and $g$ to be at least tree and two times differentiable, respectively.

The transformation corresponding to the generator $X_{\infty}$ is
\begin{equation} \label{KaptsovEI:XinfGrTransform}
\bar{t} = t, 
\qquad
\bar{x} = x + \varepsilon f,
\qquad
\bar{y} = y,
\qquad
\bar{H} = H + \varepsilon (g - y f^\prime), 
\end{equation}
where $\varepsilon$ is the group parameter.

\smallskip

Next, by the standard procedure~\cite{KaptsovEI:bk:Ovsiannikov1978}, differential invariants (up to the second order) of the generator~$X_{\infty}$ are derived. 
These invariants are as follows.
\begin{equation} \label{KaptsovEI:DiffInvs}
t,  \quad
y, \quad
H_x, \quad
H_y H_{xx} - H_{tx},\quad
H_{xx}, \quad
H_{xy}, \quad
H_{yy}.
\end{equation}
According to the theory, three invariants~(by the number of independent variables of the original equation: say, $t$, $y$, and $h=H_x$) 
are chosen as new independent variables, and the remaining four invariants are expressed as 
some functions (say, $U$, $V$, $W$, and $Z$) of them. These relations constitute the \textbf{automorphic} system
\begin{equation} \label{KaptsovEI:sysAE}
\def\arraystretch{1.75}
\begin{array}{c}
H_y H_{xx} - H_{tx} = U(t, y, h), \qquad
H_{xx} = V(t, y, h), 
\\
H_{xy} = W(t, y, h), \qquad
H_{yy} = Z(t, y, h).
\end{array}
\end{equation}
Recall~\cite{KaptsovEI:bk:Ovsiannikov1978} that any two solutions of~(\ref{KaptsovEI:sysAE}) are related 
by a transformation~(\ref{KaptsovEI:XinfGrTransform}) corresponding to the generator~$X_{\infty}$.

\smallskip

The \textbf{resolving} system represents the compatibility conditions of equation~(\ref{KaptsovEI:EqMain0}) 
and automorphic system (\ref{KaptsovEI:sysAE}) and has the following form.
\begin{equation} \label{KaptsovEI:sysRE}
\def\arraystretch{1.75}
\begin{array}{c}
    {V_y}+ {W} {V_h} - {V} {W_h} = 0,
    \qquad
    {W_y}+ {W} {W_h}- {V} {Z_h} = 0,
    \\
    {V_t}+{V}{U_h} - {U}{V_h} -{V} {W} = 0,
    \qquad
    {W_t}+{U_y}+{W}{U_h}- {U} {W_h}-{V} {Z} = 0,
    \\
    Z_t - U Z_h - V U_h + \left(W Z_h+V W_h+Z_y+\beta\right) h + V W = 0.
\end{array}
\end{equation}

In~\cite{KaptsovEI:bk:Kaptsov2023GroupFol} it was shown that system~(\ref{KaptsovEI:sysRE}) is consistent and in involution. 
It admits the tree-dimensional Lie algebra
\begin{equation} \label{KaptsovEI:algY}
\def\arraystretch{1.75}
\begin{array}{c}
\displaystyle
Y_1 = \partial_{t},
\qquad
Y_2 = \partial_{y},
\\
Y_3 = -t\partial_{t} + y \partial_{y} + 2 h \partial_{h} 
        + 3 U \partial_{U}
        + V \partial_{V}
        + W \partial_{W}
        + Z \partial_{Z},
\end{array}
\end{equation}
which is isomorphic to the subalgebra $\{ X_1, X_2, X_3 \}$ of~(\ref{KaptsovEI:alg}).
The optimal system of subalgebras of (\ref{KaptsovEI:algY}) is~\cite{KaptsovEI:bk:PateraWinternitz1977}
\begin{equation} \label{optSysY}
\{ Y_1, Y_2, Y_3 \},
\;
\{ Y_1, Y_2 \},  \; \{ Y_1, Y_3 \}, \; \{ Y_2, Y_3 \},
\;
\{ Y_1 \}, \; \{ Y_2 \}, \; \{ Y_3 \}, \; \{ Y_1 \pm Y_2 \}.
\end{equation}


Generally speaking, system~(\ref{KaptsovEI:sysRE}) is not necessary for constructing automorphic systems from exact solutions. 
However knowledge of its reductions on subalgebras~(\ref{optSysY}) proves useful from a practical standpoint.

\section{Automorphic Systems Corresponding to Exact Solutions}
\label{KaptsovEI:sec:Sec3_Results}

Analyzing the previously known exact solutions of the~GFE in~\cite{KaptsovEI:bk:Kaptsov2023GroupFol}, 
the author encountered some difficulties in constructing the automorphic systems corresponding to them, 
which motivated the consideration of the further examples.
The present section is focused on three examples of exact solutions of~(\ref{KaptsovEI:EqMain0}), the derivation of which from group 
foliations requires some additional considerations. 
In the first and second examples, knowing the invariant subalgebra of algebra~(\ref{KaptsovEI:algY}), 
to which the exact solution corresponds, proves to be beneficial. 
In the first example the final result is obtained by algebraic transformations, 
and in the second example, coordinates of polar type should be introduced.
In the third example, consideration of a specific property of the automorphic system is essential.

\subsection{Example 1: Polynomial Solution} 

Consider the polynomial solution~\cite{KaptsovEI:Bihlo_2009a}
\begin{equation} \label{KaptsovEI:sol3}
H = c_1 (x^2 - 3 y^2) x + c_2 (3 x^2 - y^2) y - \frac{\beta}{8} (x^2 + y^2) y,
\end{equation}
$c_i = \textrm{const}$, $i=1,2$, 
which corresponds to the subalgebra~$\{ X_1, X_3 \}$ of~(\ref{KaptsovEI:alg}), 
or the subalgebra~$\{ Y_1, Y_3 \}$ of~(\ref{KaptsovEI:algY}). 
The invariant of the latter subalgebra is~$\lambda = h y^{-2}$.
In this case, according to~\cite{KaptsovEI:bk:Kaptsov2023GroupFol}, the solutions of the resolving system must be of the form
\[
U = y^3 \widetilde{U}(\lambda),
\qquad
V = y \widetilde{V}(\lambda),
\qquad
W = y \widetilde{W}(\lambda),
\qquad
Z = y \widetilde{Z}(\lambda).
\]
Then, the automorphic system is reduced to
\begin{equation} \label{KaptsovEI:sysAEex2}
\def\arraystretch{1.75}
\begin{array}{c}
H_y H_{xx} - H_{tx} = y^3 \widetilde{U}(\lambda), \qquad
H_{xx} = y \widetilde{V}(\lambda), 
\\
H_{xy} = y \widetilde{W}(\lambda), \qquad
H_{yy} = y \widetilde{Z}(\lambda).
\end{array}
\end{equation}
Substituting (\ref{KaptsovEI:sol3}) into (\ref{KaptsovEI:sysAEex2}) and carrying out standard simplifications, one derives
\[
\displaystyle
\widetilde{V}(\lambda) = \frac{1}{4} \sqrt{
    192 c_1 \lambda + c_3},
\qquad
\widetilde{W}(\lambda) = \frac{24 c_2 - \beta}{24 c_1}\, \widetilde{V}(\lambda)
        - \frac{1}{96 c_1} c_3,
\qquad
\widetilde{Z}(\lambda) = -\beta - \widetilde{V}(\lambda),
\]
\[
\widetilde{U}(\lambda) = \frac{\widetilde{V}(\lambda)}{288 c_1^2} \left\{
    (24 c_2-\beta) \, \widetilde{V}^2(\lambda)
    -\frac{c_3}{2} \, \widetilde{V}(\lambda)
    + 96 c_2 (c_1^2 +c_2^2)
    - \, \frac{\beta}{16} \, (\beta^2 - 72 \beta c_2 + 2880 c_1^2 + 1728 c_2^2)
\right\},
\]
where 
\[
c_3 = 576 \, (c_1^2 + c_2^2) - 48 c_2 \beta + \beta^2.
\]

\smallskip

Thus, in this case, it is possible to establish the form of the automorphic system corresponding to solution~(\ref{KaptsovEI:sol3}), knowing the reductions~(\ref{KaptsovEI:sysAEex2}) in combination with simple algebraic transformations. 
The derived solution of the resolving equation is very close to that was obtained in~\cite{KaptsovEI:bk:Kaptsov2023GroupFol} for the subalgebra~$\{ Y_1, Y_3 \}$.

\subsection{Example 2: Harmonic Solution} 

Consider the solution~\cite{KaptsovEI:Bihlo_2009a}
\begin{equation} \label{KaptsovEI:ExactSolHarmonic}
\displaystyle
H = \frac{\beta}{2} (x^2 + y^2)^{\frac{3}{2}} \sin^3 \left(\frac{1}{3} \arctan \frac{y}{x} \right).
\end{equation}

\begin{remark}
There is also the solution~\cite{KaptsovEI:Bihlo_2009a}
\begin{equation} \label{KaptsovEI:ExactSolHarmonic2}
\displaystyle
H = -\frac{\beta}{2} (x^2 + y^2)^{\frac{3}{2}} \sin^3 \left(\frac{1}{3} \arctan \frac{y}{x} \pm \frac{\pi}{3}\right),
\end{equation}
which can be analyzed in a similar way, so we do not dwell on it here.
\end{remark}

As in the previous example, this solution corresponds to the subalgebra~$\{ Y_1, Y_3 \}$, and the automorphic system is of the form~(\ref{KaptsovEI:sysAEex2}).
However, direct calculations do not lead to clear expressions for the functions~$\widetilde{U}$, $\widetilde{V}$, $\widetilde{W}$, and~$\widetilde{Z}$.
One can overcome this issue by introducing coordinates~$r$ and~$\theta$ as follows
\begin{equation} \label{KaptsovEI:rThetaTr}
r = \sqrt{x^2 + y^2},
\qquad
\theta = \frac{1}{3} \arctan \frac{y}{x}.
\end{equation}
Then, the invariant $\lambda = h y^{-2}$ becomes
\begin{equation} \label{KaptsovEI:LambdaInv}
\lambda = \lambda(S) = \frac{\beta S (8 S^2 - 3) \sqrt{1 - S^2}}{(4 S^2 - 3)^2},
\end{equation} 
where, for brevity, we denote $S = S(\theta) = \sin \theta$.
Thus the invariant depends on only the variable~$\theta$.

Then, differentiating the invariant~$\lambda$, one derives the automorphic system in the following implicit form
\[
H_{xx}|_{(\ref{KaptsovEI:rThetaTr})} 
= \left(y \widetilde{V}(\lambda)\right)\Big|_{(\ref{KaptsovEI:rThetaTr})}
= -\frac{4 \beta r}{3}
    \left(
        8 (6 S^2 - 11) S^4 + 48 S^2 - 9
    \right) S^3,
\]
\[
H_{xy} |_{(\ref{KaptsovEI:rThetaTr})} 
= \left(y \widetilde{W}(\lambda)\right)\Big|_{(\ref{KaptsovEI:rThetaTr})} 
= -\frac{2 \beta r}{3}
    \left(
        32 (3 S^2 - 4) S^4 + 44 S^2 - 3
    \right) S^2 \sqrt{1 - S^2},
\]
\[
H_{yy} |_{(\ref{KaptsovEI:rThetaTr})} 
= \left(y \widetilde{Z}(\lambda) \right)\Big|_{(\ref{KaptsovEI:rThetaTr})} 
= \frac{\beta r}{3}
    \left(
        32 ( 6 S^2 - 11 ) S^6 
        + 24 (8 S^2 - 1) S^2
        + 1
    \right) S,
\]
\[
\left(H_{y} H_{xx} - H_{tx} \right)|_{(\ref{KaptsovEI:rThetaTr})}
= \left(y^3 \widetilde{U}(\lambda) \right)\Big|_{(\ref{KaptsovEI:rThetaTr})}
= \frac{2 \beta^2 r^3}{3}
    \left(
        48 (S^4 + 1) S^2 - 88 S^4 - 9
    \right) 
    \left(
        16 S^4 - 14 S^2 + 1
    \right)
    S^5,
\]
In order to obtain the latter relations, it is necessary to take into account the identities
\[
\displaystyle
\sin{3 \theta}  = (3 - 4 \sin^2{\theta}) \sin{\theta},
\quad
(\sin{\theta})^\prime = \sqrt{1 - \sin^2{\theta}}, 
\quad
(\sin{\theta})^{\prime\prime} = -\sin{\theta}.
\]
Solving (\ref{KaptsovEI:LambdaInv}) with respect to $S$, one can then find explicit dependencies $\widetilde{V}$, $\widetilde{W}$, $\widetilde{Z}$, and $\widetilde{U}$ of $\lambda$. E.g., the function $\widetilde{W}$ reads
\[
\widetilde{W}(\lambda) = \frac{
        2 \beta \sqrt{1-R^2} (96 R^6-128 R^4+44 R^2-3) R
    }
    {3 (4 R^2-3)},
\]
where $R=R(\lambda)$ is a solution of the equation
\[
64 (\beta^2 + 4 \lambda^2) R^8
    - (112 \beta^2 + 768 \lambda^2) R^6
    + (57 \beta^2 + 864 \lambda^2) R^4
    - \, 9 (\beta^2 + 48 \lambda^2) R^2
    + 81 \lambda^2 = 0.
\]
The latter equation can be reduced to a quartic equation by substitution $R^2 = \widetilde{R}$. 
Although solutions of the reduced equation can be obtained with the help of computational software, 
they are still quite cumbersome. As an example, here we give just one of them, namely
\begin{multline*}
\displaystyle
R_*^2(\lambda) = \widetilde{R}_*(\lambda) 
= 
q_2^{-1/6} q_1^{-1/4} (16 \beta^2+64 \lambda^2)^{-1} \,
\left\{
    \sqrt{\beta} q_1^{3/4}
    + (7 \beta^2+48 \lambda^2) \, q_1^{1/4} q_2^{1/6}
    \right. \\ \left.
    + \, \sqrt{\beta} \left[
        \left(
            (22 \beta^2 - 112 \lambda^2) \beta  q_2^{1/3}
            - (\beta^2 + 4 \lambda^2) q_2^{2/3}
            + 476 \beta^2 \lambda^2
            - 25 \beta^4
            + 2304 \lambda^4
        \right) q_1^{1/2}
        \right. \right. \\ \left. \left.
        + \, 1152 \left(
            \sqrt{\beta} \lambda^4
            + \frac{59 \beta^{5/2} \lambda^2}{48}
            + \frac{\beta^{9/2}}{36}
        \right) q_2^{1/2}
    \right]^{1/2}
\right\},
\end{multline*}
where
\[
q_1 = \beta^2 q_2^{2/3} + 11 \beta^3 q_2^{1/3} 
    + 25 \beta^4 - 2304 \lambda^4
    +(4 q_2^{2/3}-56 \beta q_2^{1/3} -476 \beta^2) \lambda^2,
\]
\[
q_2 = 14688 \beta \lambda^2 - 125 \beta^3 + 144 \lambda \sqrt{9216 \lambda^4+9204 \beta^2 \lambda^2-125 \beta^4}.
\]

\smallskip

In this example, through introducing new coordinates of polar type, it is possible 
to write down the automorphic system. 
Clearly, such a solution is extremely problematic to obtain from the resolving system directly. 
This explains the absence of invariant solutions~(\ref{KaptsovEI:ExactSolHarmonic}) and~(\ref{KaptsovEI:ExactSolHarmonic2}) in~\cite{KaptsovEI:bk:Kaptsov2023GroupFol}.

\subsection{Example 3: Solution with Constraint} 



Consider the solution~\cite{KaptsovEI:bk:HandbookLie_v2} 
\begin{equation}\label{KaptsovEI:HofQ}
H = Q(t, x).
\end{equation}
Substituting (\ref{KaptsovEI:HofQ}) into (\ref{KaptsovEI:EqMain0}), one derives the constraint
\begin{equation}  \label{KaptsovEI:constrQ}
Q_{txx} + \beta Q_x = 0.
\end{equation}
Integrating with respect to~$x$, one also finds that the solution~$Q$ satisfies the telegraph equation
\[
Q_{tx} + \beta Q = \tau(t), 
\]
where $\tau$ is an arbitrary function of $t$.

\medskip

This solution corresponds to the subalgebra~$\{ X_2 \}$ of (\ref{KaptsovEI:alg}). 
For the group foliation, this is equivalent to considering the one-dimensional subalgebra~$\{ Y_2 \}$ of~(\ref{KaptsovEI:algY}) 
with invariants~$t$ and~$h = H_x$. 

To derive solution~(\ref{KaptsovEI:HofQ}), one assumes $W=0$ and $Z=0$ (this is obviously follows from the condition~$Q_y=H_y=0$). 
Then, the automorphic system becomes
\begin{equation}\label{KaptsovEI:AEex1a}
H_{xx} = V(t, h),
\quad
H_{y} H_{xx} - H_{tx} = U(t, h),
\quad
H_{yy} = 0,
\quad
H_{xy} = 0.
\end{equation}
Integrating $H_{yy} = H_{xy} = 0$, one derives
\[
H = \widetilde{Q}(t,x) + q y,
\]
where $q=q(t)$ is an arbitrary function of $t$. 
As all solutions of the automorphic system are equivalent up to transformation~(\ref{KaptsovEI:XinfGrTransform}), one can put $q=0$.
Indeed, applying the transformation
\[
\bar{t} = t, 
\qquad
\bar{x} = x - \int \! q \, dt,
\qquad
\bar{y} = y,
\qquad
\bar{H} = H + q y, 
\]
one derives a solution of type~(\ref{KaptsovEI:HofQ}):
\[
H = \bar{Q}(t, x) = \widetilde{Q}\left(t, x  - \int \! q \, dt\right).
\]
Then, the automorphic system~(\ref{KaptsovEI:AEex1a}) is equivalent to
\begin{equation} \label{KaptsovEI:AEex1b}
H_{xx} = V,
\qquad
H_{tx} = -U,
\qquad
H_y = 0,
\end{equation}
where the functions $U$ and~$V$ must satisfy the reduced resolving system
\begin{equation} \label{KaptsovEI:REex1}
V U_h - \beta h = 0,
\qquad
V_t - U V_h + \beta h = 0.
\end{equation}
The general solution to~(\ref{KaptsovEI:REex1}) can be written in terms of quadratures of Bessel functions of the first and second kinds.

\begin{remark}
System (\ref{KaptsovEI:REex1}) can also be obtained constructively, following the definition of a resolving system, i.e. as compatibility conditions for~(\ref{KaptsovEI:constrQ}) and~(\ref{KaptsovEI:AEex1b}).
\end{remark}

\medskip

Here, to get a reduction of the group foliation corresponding to solution~(\ref{KaptsovEI:HofQ}), 
it is necessary to employ an important property of the automorphic system~(\ref{KaptsovEI:sysAE}):~all its solutions 
are related by transformation~(\ref{KaptsovEI:XinfGrTransform}). 
If one does not take advantage of this fact explicitly, then system~(\ref{KaptsovEI:AEex1b}) is derived from~(\ref{KaptsovEI:AEex1a}) 
only by artificially introducing the additional constraint~$H_y=0$.





\section{Conclusions}
\label{KaptsovEI:sec:Sec4_Concl}

Some previously known exact solutions of~(\ref{KaptsovEI:EqMain0}), obtained by standard methods alternative to the group foliations approach, are considered. 
The question arises regarding automorphic systems and particular solutions of the resolving system to which they correspond.
Addressing this question proves useful in two ways. 
Firstly, it is often unclear how the differential invariants included in an automorphic system should be related to correspond to the exact solution under consideration. Finding these relations is equivalent to determining a particular solution of the resolving system. As demonstrated in the examples, this problem is not always straightforward, and the resulting solution to the resolving system can have a rather complex structure, requiring auxiliary parametrization.

Secondly, identifying the explicit form of an automorphic system can be intriguing in itself. 
As knowing any particular solution of an automorphic system, one can find the entire set of its solutions~\cite{KaptsovEI:bk:Ovsiannikov1978}, 
reconstructing an automorphic system from a particular solution immediately leads to its integration. 
This can be especially valuable if the automorphic system is of any physical or applied interest.

For the exact solutions under consideration, automorphic systems are constructed. 
For the first example, the automorphic system can be written explicitly, 
and for the remaining examples, the automorphic systems are presented in implicit form.
These results complement those recently obtained in~\cite{KaptsovEI:bk:Kaptsov2023GroupFol}.

\section*{Acknowledgements}
This work was supported financially by the Russian Science Foundation (project code 23-11-00027). 
The author is grateful to Prof. S.~V.~Meleshko for fruitful discussions
and sincerely appreciates the hospitality of the Suranaree University of Technology.

\bigskip




\end{document}